\documentclass[12pt,dvips]{article}

\usepackage{amsmath,amssymb,exscale}
\usepackage{array,multicol}
\usepackage{afterpage,float,flafter}
\usepackage{epsfig,rotating,pifont}
\usepackage{cite}
\@addtoreset{equation}{section} \makeatother

\def\beq{\begin{eqnarray}}
\def\eeq{\end{eqnarray}}

\def\lsim{\mathrel{\rlap{\lower3pt\hbox{\hskip0pt$\sim$}}
    \raise1pt\hbox{$<$}}}         %less than or approx. symbol
\def\gsim{\mathrel{\rlap{\lower4pt\hbox{\hskip1pt$\sim$}}
    \raise1pt\hbox{$>$}}}         %greater than or approx. symbol
\usepackage{graphicx}
\usepackage{psfrag}

\title{
\vspace{-3cm}
\begin{flushright}
\small{CERN-PH-TH/2009-054\\  MPP-2009-45}
\end{flushright}
\vspace{2.7cm}
\huge{Phenomenology of $10^{32}$ Dark Sectors}
\vspace*{0.7cm}
\author{
\Large {\text{Gia Dvali$^{a,b,c}$}\footnote{georgi.dvali@cern.ch}  ~and  \text{Michele Redi$^d$}\footnote{michele.redi@epfl.ch}}\\ \\
\emph{$^a$ CERN, Theory Division, CH-1211 Geneva 23, Switzerland }\\
\emph{$^b$ CCPP, Department of Physics, New York University}\\
\emph{4 Washington Place, New York, NY 10003}\\
\emph{$^c$ Max-Planck-Institute for Physics} \\ \emph{Foehringer Ring 6,  D-80805 Muenchen,  Germany  }\\
\emph{$^d$  ITPP, EPFL, CH-1015, Lausanne, Switzerland}}}

\date{}
\begin{document}
\maketitle \thispagestyle{empty} \vspace*{-.8cm}

\begin{abstract}

We postulate an exact permutation symmetry acting on $10^{32}$
Standard Model copies as the largest possible symmetry extension of the Standard Model.
This setup automatically lowers the fundamental gravity cutoff down to
TeV, and thus, accounts for the quantum stability of the weak
scale. We study the phenomenology of this framework and show that
below TeV energies the copies are well hidden, obeying all the
existing observational bounds. Nevertheless, we identify a potential low
energy window into the hidden world, the oscillation of
the neutron into its dark copies. At the same time, proton decay
can be suppressed by gauging the diagonal baryon number of the
different copies. This framework offers an alternative approach to
several particle physics questions. For example, we suggest
a novel mechanism for generating naturally small neutrino masses
that are suppressed by the number of neutrino species.  The mirror copies of the Standard Model naturally house dark matter candidates. The general experimentally observable
prediction of this scenario is an emergence of strong gravitational effects at the LHC.
The low energy permutation symmetry powerfully constrains the form of this new
gravitational physics and allows to make observational predictions, such as,
production of micro black-holes with very peculiar properties.
\end{abstract}

\newpage
\renewcommand{\thepage}{\arabic{page}}
\setcounter{page}{1}

\section{Introduction}

Large numbers have always fascinated physicists at least since
Dirac times \cite{dirac}. One incarnation is the hierarchy problem
of particle physics, the huge ratio between the weak scale
(TeV) and the Planck scale, $M_p^2/TeV^2=10^{32}$. Within the
Standard Model (SM) this ratio is set by the VEV of the Higgs
field which is unnatural due to the quadratic divergences in the
Higgs mass. The hierarchy problem is therefore not about the big
number {\it per se}, but about its inexplicable stability in the
light of its quadratic sensitivity towards the ultraviolet cutoff
of the theory.

A radical approach to the problem is to assume that the
fundamental scale of gravity is TeV and that $M_p$ is a derived
scale. In this way the hierarchy problem is erased because the
quadratic divergences to the Higgs mass are cut-off at TeV, making
its value around the weak scale natural. This point of view,
originally introduced in the large extra dimensions scenario
\cite{ADD1}, has been recently revived in the framework of theories
with large number of particles species \cite{bound,us}.
There it was shown that  in \emph{any} theory with $N$ particle species,
\begin{equation}
M_p^2\ge N M_*^2 \, ,\label{bound}
\end{equation}
where $M_*$ is the fundamental scale of gravity.  This bound follows from the consistency of large distance black hole physics \cite{bound, us, dieter, micro, quantumN}.  Notice that the simple naturalness arguments based on perturbation theory
point in the same direction\cite{greg, veneziano}.  Interestingly, the same bound on
$N$ also holds in theories where $N$ labels elements of a discrete
symmetry group, e.g., $Z_N$ \cite{bound,discrete} suggesting that
both large symmetries as well as many species are incompatible
with a high cutoff. In the case of many species the bound
implies that $N\,  \sim \, 10^{32}$ elementary particle species
automatically lower the cutoff to $M_* \, \sim $TeV, thus
explaining  the quantum stability of the hierarchy between the weak and the Planck scales.
The important point is that the above bound holds independently of
the nature of the particles. For example the large extra dimensions proposal
automatically falls in this class.  This is because the relation
(\ref{bound}) can be viewed as the relation between the
higher dimensional ($M_*$) and  four-dimensional Planck scales,
where $N$ is the number of Kaluza-Klein (KK) modes. Indeed, the
volume of the compactified extra space, measured in units of the
fundamental Planck length $1/M_*$, counts the number of  KK
species, irrespectively of the precise shape of the manifold or
the number of dimensions. Thus, whenever $M_*\, \sim $TeV, from a
4D point of view there are $10^{32}$ Kaluza-Klein resonances below
the fundamental gravity scale.

In this paper, we shall focus on another extreme case, in which
the species are not KK states, but identical copies of the SM.
The idea of identical copies of the SM was already introduced in \cite{kaloper,foot2} in different
contexts. In the case of two copies parity symmetric worlds this was also considered in \cite{foot1}
(see also \cite{mirror1} for earlier discussions). In our case we postulate
the existence of $10^{32}$ mirror copies of the SM coupled only through gravity as
this would automatically explain the hierarchy. At this
level of the discussion, the large number $N\,= \,10^{32}$ is an
UV-insensitive input, which is shared by other approaches to the
hierarchy problem. For example, in the simplest realizations of the sofly broken
supersymmetric SM \cite{susy} an equally large input number is the
ratio between the Planck mass and the soft supersymmetry-breaking
masses.

However, in the present case there is a new rationale for the emergence of a
large number $N \sim 10^{32}$  from the symmetry requirement. Neglecting the hierarchy
problem, let us ask the following question. What is the largest possible exact symmetry
extension of the SM group under which the SM particles transform non-trivially?
Existence of a continuous symmetry is excluded for the
following reasons.  First, the non-observation of the corresponding massless gauge bosons,
immediately rules out continuous gauge symmetries, unless the corresponding gauge
coupling is extraordinarily small.  For example, already the bounds on long-distance gravity-competing forces (see \cite{equiv}),
constrain the strength of the new gauge forces at most to $\alpha_{new} \,  \sim \, 10^{-50}$ or so.
In such a case, the symmetry in question for all practical purposes is essentially a global symmetry which however is forbidden in quantum gravity theories. In fact it was shown in \cite{discrete} that from
the consistency of the BH physics, any continuous global symmetry must be at most
broken down to a discrete subgroup with maximum  $N \equiv M_P^2/M_*^2$ commuting elements, where as before  $M_*$ is the fundamental scale where gravity gets strong.
Since, irrespectively of the hierarchy problem, the ultimate phenomenological bound on $M_*$ is around TeV, we automatically get  the upper bound on $N$ being $10^{32}$ or so. From purely large distance considerations, we
are then left with discrete symmetries as  possible unbroken
extensions of the SM group.  The largest  of these is
permutation symmetry acting on $10^{32}$ SM copies.
Thus, the  requirement of a phenomenologically acceptable maximal exact
symmetry, automatically solves the hierarchy problem. Another motivation
is provided by the strong CP problem \cite{glennys}.

In this paper we shall investigate several phenomenological
aspects of the above proposal. The physics of our interest comes from
two types of considerations. First,  restricting the strength of
all possible low energy interactions among the SM copies solely by
symmetries and consistency requirements, such as unitarity, we
uncover new phenomenologically interesting phenomena.  One result
is a novel mechanism of small neutrino mass generation, which
comes out to be $1/\sqrt{N}$-suppressed in our scenario.
Another interesting effect is a potentially-observable oscillation of neutron into its
hidden copies. The novelty in comparison to previous studies lies in  the large number of
``neighboring''  copies into which neutron can oscillate.
Secondly,  we discuss the phenomenology of  the short distance
completion of the gravitational sector that is imposed upon us by
consistency. In this regard our approach is different from the large
extra dimensional scenario, in which geometry is an input. In the
present case,  the underlying gravitational physics is an outcome
of consistency and of the well-established  large distance
dynamics, such as thermodynamics of macroscopic black holes.
Inevitability of the new gravitational dynamics is revealed  when
the above well-known large distance properties are combined with
the field theoretic consistency requirements, such as unitarity.
By supplementing  the above knowledge with symmetry,
one can go surprisingly far in understanding the properties of the
new short-distance gravitational physics.

One model independent aspect of  the present
framework is that existence of low energy species generically implies
extra-dimension-type modification of gravity at distances
parametrically larger  than  the cutoff length $l_* \, \equiv \,
M_*^{-1}$ \cite{micro}. This modification is not necessarily reducible to a
smooth geometry,  however it does exhibit two important
characteristics of extra dimensions: emergent locality in the space of species,
and inevitability of a distance scale $R\, \lesssim  l_*$ below which gravity changes classically.
The scale $R$ plays a role qualitatively similar to the radius of
the extra dimensions, beyond which the gravitational force law
gets modified. An interesting fact is that the
correspondence to classical geometry-type description can
be classified in terms of the subgroups of the full permutation
group.  In particular, the  full permutation symmetry $P(N)$ is associated to
the maximal possible departure from an extra dimensional picture.

The connection between symmetry and geometry arises through black-hole (BH) physics.
An essential feature of the framework, hinging solely on
unitarity, is that small BHs cannot decay democratically into all
the species but predominantly into the specie that produced them
in the collision process. This feature is in stark contrast with
the standard macroscopic BH physics, but is characteristic to the
small BHs in the large extra dimensions scenario. In the
latter case, a BH that is smaller than the size of  the
extra dimensions predominantly decays in the four-dimensional
species that are localized within its reach in the extra space,
while it cannot decay into the distant ones (this phenomenon and its
reconciliation with the known universal thermal properties of the
quasi-classical black holes was discussed in detail
in \cite{oriol}). As we increase the mass of the BH more
species become available for the BH evaporation and eventually
large BHs decay universally (thermally) into all the species as
dictated by semi-classical gravity. This can only happen for BHs
whose mass is larger than $M_P \sqrt{N}$ because, as was shown in
\cite{us,micro,oriol} below this mass BHs never behave as four
dimensional Einsteinian (classical) BHs. Thus, from the point of
view of the quasi-classical BH physics and gravity, species behave
as if they are indeed separated by an extra dimension. That is,
the BH evaporation allows to define a metric in space of species:
the distance between specie $i$ and $j$ can be related to the
critical size of the BH of specie $i$ to decay into particles from
specie $j$. In other words,
\begin{equation}
{\rm UNITARITY + BLACK~ HOLE~CONSISTENCY   \rightarrow LOCAL~GEOMETRY}\,.
\end{equation}
Depending on the degree of symmetry the geometrical description can be
in fact real \cite{micro}. For example requiring cyclic symmetry
between the species, the gravitational dynamics becomes
describable  in terms of  classical extra-dimension of size
$R>>1/M_*$.

The maximal departure from the smooth geometric interpretation  is
obtained considering the limit in which all the $10^{32}$
copies of the SM are related by an exact permutation symmetry.  In
this limit any point in the space of species becomes ``equidistant'' from any other.
Correspondingly, the phenomenology of the gravitational sector of
the cyclically symmetric extension of the SM shares some
qualitative features with the one of the large extra dimensions
\cite{ADD3}, but there are crucial differences, for example the evaporation
rate of the microscopic BHs. The case of full permutation symmetry on the
other hand presents rather different and novel features.

Having removed the crutches of the smooth geometry, the
permutation symmetric SM becomes very  predictive since the
couplings are highly constrained by the full permutation symmetry.
This is in contrast with geometrical models where the details of the
geometry (such as number of dimensions, warping, location of the
branes etc.) are a crucial input. Unitarity requires interspecies
coupling to be strongly suppressed. This allows to automatically
satisfy all astrophysical and cosmological constraints. Many
mechanisms introduced in the context of large extra dimensions
find the corresponding implementations here,  leading, for
example, to interesting scenarios for dark matter (see the
complementary work \cite{Dvali:2009fw}) and neutrino masses.

The paper is organized as follows. In section 2 we consider the
connection between black-holes and geometry emphasizing the role
of symmetries in the space of species. In the rest of the paper we
focus on the phenomenology of the permutation symmetric SM. In
section 3 we consider astrophysical, cosmological and laboratory
bounds on the model. The possibility of mixing
between neutrons from different copies is discussed in section 4. In section 5 we present
possible mechanisms for neutrino masses and dark matter. We conclude in section 6.

\section{Black Holes and Geometry in the Space of Species}
\label{geometry}

In this section we wish to recall the close relationship between
geometry and BH evaporation.
%The reader only interested in the phenomenological
%implications of the permutations symmetric scenario should directly skip to
%Section 3.
Material related to the one here appeared in \cite{us,micro}
and further properties of small black holes in theories with large
number of species are discussed in \cite{oriol}.

We will consider a theory space with $10^{32}$ exact copies of the
SM coupled to gravity. According to eq. (\ref{bound}) the presence
of the $N$ species lowers the quantum gravity scale at $M_*\sim
M_p/\sqrt{N}\sim$ TeV. Since gravity becomes strong around TeV,
the model indepedent prediction of this framework is the existence
of microscopic black holes with mass $M_{BH} \, \geq \, M_*$. Such
micro-BHs would be produced in particle collisions at
energies above $M_*$. Thus, if the hierarchy problem is solved by the
low quantum gravity scale, this will be directly probed at the LHC.

The existence of large number of species in the low energy theory, dramatically
affects the physical properties of the micro-BHs.
The absolute lower bound on the BH mass, below which these objects can no longer
be treated as normal Schwarzschild BHs,  is given by \cite{micro},
\begin{equation}
\label{BHcros}
M_{Schwarzschild~BH} \, \geq \, M_P\, \sqrt{N} \,.
\end{equation}
For this value of the mass the curvature at the horizon of a
classical BH is of the order of $M_*$ so BHs lighter than $M_P
\,\sqrt{N}$ must be non-Einsteinian. The departure from the
Einsteinian regime can however start at even larger distances
already at the classical level and delay the quantum regime up to
energies $\sim M_*$. As we shall see on the example of our present framework,
this feature is model dependent and closely related to symmetry properties of the theory.

The other major difference concerns the properties of the BH ``hair''.
It is well known that in Einsteinian gravity classical BH satisfy
no-hair theorems\cite{nohair}. The absence of hairs implies that
BHs can only be labelled by charges that can be measured at
infinity, either classically or quantum mechanically. As a result,
in the absence of such charges, the Hawking evaporation process of
a large Schwarzschild black hole is thermal and is completely
democratic between the species. In the evaporation process of a BH
of temperature $T_H$ all the thermally accessible species (with
masses $m \, \lsim \,  T_H$) will be produced at the same rate
$\Gamma \sim T_H$, whereas the production of the heavy species ($m
\, \gg \, T_H$) will be Boltzmann suppressed by the factor
$e^{-m/T}$.

However,  the nature of the micro black holes in theories with large
number of fields is dramatically different from their macro counterparts \cite{us,micro, oriol}. Unitarity implies that the decay of the smallest BHs is maximally
asymmetric in species. In order to see this, let us follow the argument of \cite{us} and  consider a  production of a lightest micro BH in the collision of particles and anti-particles
in $i$-th  SM copy. Such a BH will be produced with probability of order one at a
center of mass energy $\sim M_*$, and will have a characteristic
mass $M_{BH} \sim M_*$. As a reference to the SM copy of
origin, we shall endow such a BH by an index $i$.  This process then can expressed as,
\begin{equation}
\label{BHi}
\Phi_i \, + \, \bar{\Phi}_i \, \rightarrow \, {\rm BH}_i \, .
\end{equation}
For a macroscopic Schwarzschildian BH such a label would be
redundant,  because of the absence of  hair,
but for the microscopic ones it is not, as it is evident from the following reasoning. For the center of mass energy $\sim M_*$ the only scale in the problem is $M_*$,
and the production rate of the lightest BH is $\Gamma \sim M_*$.
Since the BH$_i$ was produced in particle-anti-particle collision
in the $i$-th SM copy, it does not carry any internal quantum
number, and by symmetries alone could easily decay into
particle-anti-particle pairs of any other $j$-th copy. For a
classical Einsteinian BH this decay rate would indeed be
$j$-independent. However, for the microscopic BH in question, this
is impossible. By CPT invariance, the BH$_i$ should be able to
decay back into a pair of $i$-th species with the rate
$\Gamma_{ii} \sim M_*$. But then by unitarity,  it cannot decay
into all $N$ other individual copies of SM with  the same partial
rates.  The decay rate into the majority of the SM copies must be
$\sim M_*/ N$, or else unitarity would be violated much below
$M_*$ energies.

In the limit of exact permutation symmetry, the BH$_i$ decay rates
are extremely restricted, since decay rates into all $j \neq i$
copies must be strictly equal, for any values of $i,j$.  That is,
the decay rates split into the diagonal and non-diagonal ones,
\begin{equation}
\label{rates} \Gamma_{ii} \, = \,
\Gamma_{diag}(M_{BH}),~~~\Gamma_{ij} \, =\,
\Gamma_{nondiag}(M_{BH}).
\end{equation}
where both diagonal and non-diagonal rates are the functions of
the BH mass. So far, what we know from unitarity is that, for
$M_{BH} \, \sim M_*$,
\begin{equation}
\label{ratesmstar}
\Gamma_{diag}(M_*) \,  \sim \, M_*,~~~\Gamma_{nondiag}(M_*) \, \lesssim \, {M_* \over N}
\end{equation}
As we will see, this suppression of the interspecies couplings (induced by BHs or not)
is the crucial feature of the permutations symmetric scenario that makes its phenomenology rather constrained.

Notice that had we limited the symmetry group to the cyclic
permutations of the SM copies, the decay rates $\Gamma_{ij}$ could have been
non-trivial periodic  functions of $i,j$.  The general unitarity requirement is
\begin{equation}
\sum_j \Gamma_{ij} (M_*) \, \lsim \, M_*
\end{equation}
In such a case one could introduce a notion of nearest neighbor(s)
in the space of species, for example,\begin{equation}
\Gamma_{ij}(M_*) \, \propto  \, M_* \, {\rm e}^{-N{\rm
sin}\left(2\pi{|i-j| \over N}\right )} \, .
\end{equation}
This is the case which naturally admits an extra dimensional
interpretation where the copies of the SM are localized on the branes
uniformly spaced around an extra dimension. Consider a small BH produced in particle collision on
$i$-th brane. If the high-dimensional gravitational radius is
smaller than the inter-brane distance,  the  BH$_{i}$ cannot
evaporate into a SM copy that is localized on a distant brane. In
this example, the non-democracy has a clear geometric meaning, and
is a consequence of locality in the extra space. In contrast, the
case of the full permutation symmetry is an extreme case, which
maximally departs from any conventional geometric picture, since
there is no notion of a nearest neighbor there.

\subsection{The Classical Crossover Length Scale}

As shown in \cite{micro}, the non-democratic evaporation of the
smallest BHs, inevitably leads to the existence of the second
length  scale, which we shall denote by $R$. The key point is that
$R$ is larger or of the order of the fundamental Planck length
$M_*^{-1}$, and marks the crossover distance, beyond which the BHs
become normal Einsteinian  BHs. That is, for BHs larger than $R$
the index $j$ becomes redundant and their evaporation becomes
democratic in all the species.

At this point the BH horizon is related to the mass of the BH via the usual
Schwarzschild relation
\begin{equation}
R \, =  \, 2M_{BH} \, G_{Newton}.
\label{R}
\end{equation}
At the intermediate distances
\begin{equation}
 M_*^{-1} \, \ll \, r \,  \ll \, R \, ,
\end{equation}
gravity is still classical, but the micro BHs of  gravitational
radius within this interval cannot be Einsteinian BH. In
particular in their evaporation  spectrum there is an
$r_g$-dependent bias with respect to some species. The black holes
of sizes $r_g \, \sim \, M_*^{-1}$ and $r_g \, \gsim \, R$ are
extreme cases, corresponding to the  maximally biased  and
unbiased cases.

In the case of $N$ standard model copies, we can give to the
distance $R$ a simple geometric meaning if we think of a label $j$
as of the coordinate of the  localization site in the extra
dimension of radius  $R$. This setup would reproduce all the
general properties of the microscopic BHs that were displayed
above. For example, BHs of gravitational radius smaller that the
size of the extra dimension, but still larger than the fundamental
scale, will be classical objects. However, their evaporation
process will be undemocratic in copies of SM since the small BH
will not be able to reach out to all the localization sites. On
the other hand BHs larger than the size of extra dimension would
be normal, and thus completely democratic with respect to all the
species. It is however important to stress that the interpretation
of the underlying gravitational dynamics in terms of  the
conventional extra dimensions may not be possible in general. In
particular this is the case if we require the exact full
permutation symmetry between the SM copies.  This setup
corresponds to the maximal departure from a smooth geometry.

To investigate the properties of microscopic BHs in our
case we can follow the general strategy in \cite{micro},
focusing on the theory with $N$ exact SM copies. We require that
the physics, including the gravitational one, is identical as seen from each copy.  In
symmetry terms, this can be guaranteed for instance by requiring
the cyclic permutation symmetry under which
$j$-th  SM copy is replaced by $j-1$-th one (this operation,  of
course,  requires an obvious periodicity in numbering $ j \,
\equiv \, j + N$). This symmetry, still leaves room for further
restriction, since we may or may not require the symmetry under
full permutation group.  We keep such an option open at the
moment.

Since at  distances $ r \, < \, R$ gravitational interaction among
the different  SM copies is no longer universal,  we should keep
the  labels explicit.  Consider a Newtonian interaction between
the two point-like sources, belonging to the $i$-th  SM copy, of
masses $M_i$ and $m_i$, and assume the separation by a distance
$r$. Following \cite{micro},  this interaction can be
parameterized by the following gravitational potential (below
everywhere we shall ignore numbers of order one)
\begin{equation}
\label{v}
V(r) \, = \, {M_im_i \over M_*^2} {1 \over  r  \, \nu (M_*r)} \, ,
\end{equation}
where $\nu(M_*r)$ is  some smooth function,  such that
\begin{equation}
\label{nu}
\nu \sim 1~~{\rm for} ~~ r\sim M_*^{-1},~~ {\rm and}~~
\nu \simeq  {M_P^2 \over M_*^2} ~~{\rm for}~~ r \, > \, R \, .
\end{equation}
The latter boundary conditions follow  from the fact that by
changing $r$  from $R$ to  $M_*^{-1}$, the above interaction must
interpolate  between the usual Newtonian force and the strong
gravity.

Since at the intermediate distances gravity is weakly coupled,
the effective gravitational radius ($r_g^{ii}$)  of the source of
mass $M_{BHi}$ seen by the particles of the same $i$-th species
can be estimated from the equation,
\begin{equation}
\label{v}
1\, = \,  {M_{BHi} \over M_*^2} {1 \over  r_g^{ii} \, \nu (M_*r_g^{ii})} \, ,
\end{equation}
that is from the condition that the Newtonian potential becomes
order one. Although, $r_g^{ii}$ represents  the BH horizon for the
same species, $r_g^{ii}$  is not necessarily a horizon for all the
other species, since only the species that can be produced in the
evaporation process of a given BH, can see its horizon.

For any BH belonging to $i$-th SM copy  the number of species to
which it can evaporate  is a function of the black hole mass
($r_g^{ii}$), $\mathcal{N}(M_*r_g^{ii})$.   Because of the
symmetry, this function is the same for all the copies, and its
boundary properties are similar to (\ref{nu})
\begin{equation}
\label{n}
\mathcal{N}(1) \sim 1, ~~~ \mathcal{N}(M_*R)\,  = \,  {M_P^2 \over M_*^2}
\end{equation}

In contrast the  gravitational potential between two sources of
masses $M_i$ and $M_j$  belonging to two different $i$-th and
$j$-th  SM copies, will be set by the index-dependent  function
$\nu(M_*r)_{ij}$, with the  boundary condition $\nu(MR)_{ij} =
M_P^2/M_*^2 = N$, which follows from the definition of $R$.  The
necessary condition for the particles of $j$-th  copy to be
produced in the evaporation process of the BH of mass $M_{BHi}$
made out of the $i$-th specie, is that the $j$-particles see the
horizon of the $i$-th BH. The corresponding horizon we shall call
$r_g^{ij}$.  Generalizing the notion of $r_g^{ii}$,   the
non-diagonal gravitational radius can be estimated from
\begin{equation}
\label{evap}
{M_{BHi} \over M_*^2} {1\over r_g^{ij} \, \nu(M_*r_g^{ij})_{ij}} \, = \, 1 \, .
\end{equation}
Then $r_g^{ij}$ represents the horizon of $i$-th BH with respect
to the particles from the $j$-th species.   Thus, the function
$\mathcal{N}$ counts the solutions of (\ref{evap}).

In the above ``geometric'' example, in which SM copies are
displaced in the extra dimensional space,  the non-universality of
the function $\nu_{ij}$ is  explicit and follows from the locality
in the extra space.  The important point however is that locality
in the space of species emerges automatically regardless of any
input geometrical assumptions. The resulting properties of the
space of species, can be parameterized by an effective ``metric''
in this space. Depending on the level of the inter-copies
symmetry, this metric can exhibit different level of
correspondence with the one of conventional compact dimensions. In
the extreme case of full perturbation symmetry, the departure from
any resemblance of the classical geometry will be  maximal.

\subsection{Cyclic Symmetry of the Standard Model Copies}

So far, our only symmetry requirement was,  that the laws of physics, as seen by an observers from  any SM copy,  must be identical. This requirement guarantees
that the  quantities $r_g^{ii}$,  $\nu$ and $\mathcal{N}$ are
independent of $i$.  The analog of flat geometry in the space of
species would then correspond to the following approximate form of
the  functions $\nu$ and $\mathcal{N}$,
\begin{equation}
\label{nuandn} \nu \, = \, \mathcal{N} \, = \, \left( {r_g \over R}
\right )^n {M_P^2 \over M_*^2} \, ,
\end{equation}
where $n$ is an arbitrary number. The boundary condition
$\mathcal{N}(1) = 1$ fixes $R \, = \, (M_P^2/M_*^2)^{{1\over n}}
M_*^{-1}$, and thus,
\begin{equation}
\label{nuandn1}
\nu \, = \, \mathcal{N} \, =\, (r_g M_*)^{n}\,.
\end{equation}
Can this form be guaranteed by some discrete symmetry among the SM
copies? For $n=1$, this form is indeed guaranteed by the cyclic
symmetry under which $i \rightarrow  i + 1$ for any $i\neq N$ and
$(i = N) \,  \rightarrow \, ( i = 1)$. We can then generalize to
the  required symmetry group for arbitrary $n$ in the following
way.  Let us label copies by a group of $n$ indexes as follows,
\begin{equation}
\Phi_{i_1i_2...i_n} \, ,
\end{equation}
where each index takes the value $i_{\alpha} \, = \,
1,2,...N_{\alpha}$,  with $\alpha \, = \, 1,2,...n$. Since
the total number of copies is $N$, the numbers $N_{\alpha}$ must satisfy
\begin{equation}
\prod_{\alpha} \, N_{\alpha} \, = \, N \,.
\end{equation}
We then require invariance under  permutations of copies that
are obtained by independent cyclic permutations of indexes within
each group  $i_{\alpha}$,
\begin{equation}
\label{group}
i_{\alpha} \, \rightarrow \, i_{\alpha} \, + \, 1  ~~{\rm for~any}~ i_{\alpha}\, , ~{\rm and}~
(i_{\alpha}=N_{\alpha}) \, \rightarrow \, (i_{\alpha} \, = \, 1) \, .
\end{equation}
For all $N_{\alpha}$ being equal, this symmetry uniquely fixes the
form  of the functions $\nu$ and $\mathcal{N}$ to be  given by
(\ref{nuandn1}).  This is remarkable, since a simple requirement
of  symmetry under cyclic permutation of the SM copies,  forces
the short-distance gravity to behave as if $n-$extra dimensions
with the sizes $R \, = \, M_*^{-1} \sqrt[n]{N}$ open up, although
nothing like this has ever been postulated!  Making $N_{\alpha}$-s
arbitrary, simply changes the sizes of this dimensions to
$R_{\alpha} \, = \, M_*^{-1} N_{\alpha}$.

\subsection{Full Permutation Symmetry of the Standard Model Copies}

We are now ready to consider the case which corresponds to a
maximal departure from the classical geometry, by postulating the
full permutation symmetry group of the standard model copies. That
is, we require physics to  be invariant under the  exchange of
arbitrary copies. This requirement immediately fixes all the
non-diagonal  quantities to have the same value $\nu_{ij} \, = \,
\nu'$. The consequences for BH evaporation are pretty dramatic,
since all non-diagonal decay channels must also have the same
rate, by symmetry. In this case formally the crossover scale
$R\sim M_*$ and gravity jumps directly from the classical 4D
description to the full quantum gravity regime at that scale.

\begin{figure}[htb]
\begin{center}
\includegraphics[height=5.5cm,width=6.5cm,clip]{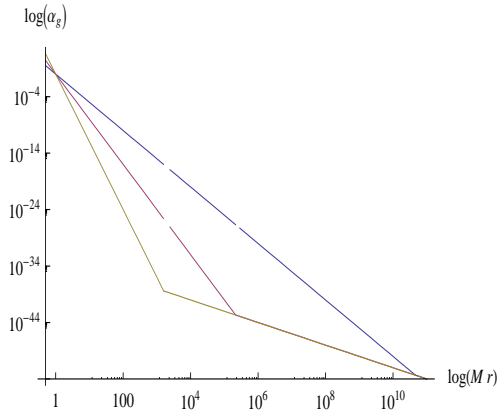}
\caption [Fig 1] {\it Gravitational force of two point particles with mass $M_*$
as function of distance with 3 (blue), 6 (red), 10 (green) extra-dimensions. The limit $n\to \infty$ formally resembles the permutation symmetric scenario.}
\end{center}
\label{strenghtfig}
\end{figure}

To such inter-species relations it is hard to give any sensible
geometric meaning because of the following
reason.  The permutation symmetry group implies that, if there is
any notion of metric in the space of species, the copies must be
equidistant in this space.  This is impossible unless the space of
species has effectively $n = N-1$ dimensionality! The only
geometric space that could imitate such an extreme democracy is
the $N-1$-dimensional space (see Fig. \ref{strenghtfig}),
however this precisely the case where any geometrical notion is lost since only the first KK modes of the
theory would be well described by the effective theory.

The fact above can also be understood from the observation that
the full permutation symmetry group can be obtained as the limit
of the above-introduced $n$-cyclic permutations for which $n \, =
\ N-1$. Indeed, in such a limit the number of indexes $i_{\alpha}$
becomes $N-1$, and each index can take only two possible values,
which can be taken to be $0,1$.  Moreover, in each sequence there
can only be a single  distinct index, e.g.,
$\Phi_{00001000,...000}$. The species then are simply identified
by the position of $1$ in this sequence.  Making the cyclic
permutation in any $i_{\alpha}$ then simply reduces to changing $0
\rightarrow 1$ or vice versa at that position. But because $1$ can
only appear once in the whole sequence, we have to make the
opposite flip in one of the other indexes. This effectively is
equivalent to  permuting a single pair of SM copies,  that had
index $1$ in these two locations.

\subsection{Democratization of Micro Black Holes}

Before concluding this section let us discuss an important property of
microscopic  BHs that reconciles their non-universality  with respect to different species
with the known thermal and democratic properties of the classical BHs.
Indeed, at first glance, such a non-universality of BH evaporation looks
puzzling, because of the following reason.  On one hand, the BHs in question are
quasi-classical thermal objects and thus are expected to radiate all the light species universally. On the other hand this is impossible by unitarity, and their non-democratic evaporation inevitably suggests that different species see different horizons. How can the two seemingly inter-exclusive properties of thermality and non-universality be reconciled?

Both of the above properties seem to be explicitly supported by the extra dimensional considerations, in which even quasi-classical  BHs that are smaller than the compactification radius must evaporate non-democratically
in various four-dimensional species simply by locality in extra dimensions.
Nevertheless,  such a situation would be problematic, since it would imply that BHs can be labeled by  their belongness to a particular species. This label will not be associated with any exactly-conserved quantum number measurable at large distances by a four-dimensional observer, in contrast to known no-hair properties\cite{nohair}.

The resolution of the puzzle was given in \cite{oriol}.  The outcome of these studies suggests that microscopic BHs on top of the quantum Hawking evaporation time, posses another intrinsically classical time scale,  a ``democratization'' time.  During the latter time any given non-democratic BH is classically unstable, and evolves in time until it becomes fully democratic in species. Thus, the non-democracy is never a property of  a classically stable neutral BH. The balance between the two time scales depends on the BH mass,
the number of species and geometry in the species space.  However,  for the smallest  BHs evaporation always wins and they never reach a fully democratic state.   For such BHs the democratization process can be ignored
and all the analysis given above is fully applicable. Applying this consideration to our present context, the importance of the democratic transition will depend on the subgroups of $P(N)$.

For example, for  the full $P(N)$-symmetry  case, the BHs that can be potentially observed at LHC will evaporate way before the democratization time.  Thus,  for such light BHs, the  democratization  time scale is irrelevant. However, for the cyclic symmetric
case, the existence of the neighboring copies gives important correction even in the dynamics of the relatively light BHs.  First, the existence of extra copies,  should  increase the evaporation rate
substantially\cite{micro},
 \begin{equation}
\label{masschange2}
{dM_{BH} \over dt} \, = \, M_*^2 \, \left ( {M_* \over M_{BH}} \right )^{{2-n \over 1+n}},
\end{equation}
which for the BH lifetime gives
\begin{equation}
\label{lifetime}
\tau_{BH} \, \simeq \, M_*^{-1} \left ({M_{BH} \over M_*}\right ) ^{{3 \over n +1}} { n+1 \over 3} \,.
\end{equation}
The above  lifetime is by a factor $\left ({M_{BH} \over M_*}\right )^{{n \over n+1}}$ shorter than the lifetime of a microscopic BH localized within $n$ flat extra dimensions.
The picture instead is such,  as if the BH is localized in $n$ flat extra dimensions in which
there are $N$ 4-dimensional light species localized at
uniformly distributed sites (3-branes).  With the growing mass, the BH horizon grows in the space of species according to $n+4~$- dimensional law,  and captures more and more sites.

Correspondingly,  the  branching ratio of BH evaporation also changes.
For a BH of some mass  $M_{BH} \, \gg \, M_*$,  produced by particles of $i$-th copy, there are  $\mathcal{N} \, = \, \left ({M_{BH} \over M_*}\right ) ^{{n \over n+1}}$  invisible
decay channels. Thus, for such a BH only a fraction
\begin{equation}
\label{fraction}
{E_{BH \rightarrow i-{\rm th~copy}} \over  E_{BH \rightarrow  {\rm all~copies}}} \sim \left ({M_* \over M_{BH}}\right )^{{n\over n+1}}
\end{equation}
of the total energy will be released back in $i$-th species. The rest will be distributed over  the extra copies.

Secondly,  the (partial) democratization process for such BHs will also be important and be comparable with the evaporation rate. In the other words,  the number of available evaporation channels can change substantially within the BH lifetime, and this in return will decrease the lifetime even further.
Thus, the micro BHs in the theory with many SM copies, posses the observable properties very distinct from the standard large  extra dimensional case,  and these differences  can potentially be tested at LHC.

\section{Bounds}
\label{bounds}

In this section we study phenomenological bounds on our theory
with $N=10^{32}$ identical SM copies. As we will see these are
very mild and never worse than in the large extra dimensions
scenario. We focus on the maximally symmetric version of the model
by requiring  invariance under the full permutation group $P(N)$.
The phenomenological bounds come from two sources:  1) First,  the new
effective interactions generated among the SM particles that are
subject to the  precision electro-weak constraints;  2) secondly, from
the production of new states in the  collision processes of the SM
particles. The role of the new states, that are either  produced
in SM collisions or are mediators of the  new interactions, can be
played either by the hidden SM copies or by new gravitational
degrees of freedom that modify gravity  at the scale $M_*$.

\subsection{Graviton Induced Processes}

\begin{figure}[htb]
\begin{center}
\psfrag{M}[][]{\Large $\frac 1 {M_p}$}
\psfrag{ep}[][]{\Large $e^+$}
\psfrag{em}[][]{\Large $e^-$}
\psfrag{j}[][]{\Large $j$}
\includegraphics[height=4.0cm,width=6cm,clip]{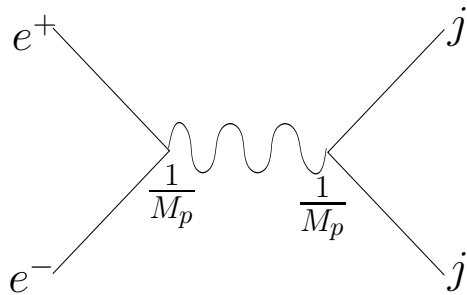}
\caption [Fig 1] {\it Interactions between visible and hidden sector mediated through graviton exchange.}
\end{center}
\label{cross}
\end{figure}

We start considering processes induced by the standard massless
graviton, which is the only field mediating interactions among the
different species at low energies. Phenomenologically,  the
crucial difference with large extra-dimensions is due to the fact
that the hidden SM copies do not couple directly to our SM fields,
but only through the graviton exchange as in Fig. \ref{cross}. On
the other hand the  KK gravitons of the large extra dimensions do couple directly to
the SM particles at tree level with $1/M_P$ strength.
The cross section for the production of the other species in the
collision of SM states due to graviton exchange is enormously
suppressed. By simple dimensional analysis,  we can display the
relative scalings of the latter cross section versus the
individual KK production in the LED case,  as follows,
\begin{eqnarray}
\sigma_{e^+e^-\to j j}&\sim& \frac {E^2}{M_P^4}~~~~~~~~~~~~~~10^{32}~SM~copies \nonumber\\
\sigma_{e^+e^-\to KK+\gamma}&\sim& \frac {\alpha}{M_P^2}~~~~~~~~~~~Large~Extra-D \, .
\end{eqnarray}
This fact ameliorates or completely removes the cosmological
constraints on the processes mediated by the graviton exchange,
despite the fact that each sector contains massless photons and
other light  states. For example as in every theory with extra
light particles, one immediate worry is that if these states were
produced in the early Universe they would destroy the successes of
standard cosmology. In particular, since Big Bang Nucleo-synthesis
(BBN) predicts the right abundances of light elements we certainly
do not want to change the cosmology at least until BBN, which
corresponds to temperatures $\approx 1 MeV$. This requires that at
most one extra relativistic degree of freedom contributes to the
Hubble expansion in the early Universe during BBN.

We can easily see that this condition is automatically satisfied
by the above graviton-mediated interactions. Let us estimate the
critical temperature ($T_C$) above which the cosmological
evolution  would radically deviate from the standard one,  due to
the production of the extra species  (in \cite{ADD3} this
temperature was called the {\it normalcy} temperature). For this,
we assume that at some initial state Universe is populated only
with our SM particles at temperature $T$. In the radiation
dominated phase the expansion rate is given by,
\begin{equation}
H\approx \sqrt{g^*} \frac {T^2}{M_P} \, ,
\label{hubblerate}
\end{equation}
where $g^*$ is the number of active degrees of freedom which  is of
order one, since by assumption only the SM particles are initially in thermal equilibrium.

If the production of states from other sectors arises
through the graviton exchange diagrams such as in Fig.
\ref{cross},  the total rate is roughly,
\begin{equation}
\Gamma_{TOT} \, \sim \,  \frac {T^5}{M_P^4} N \, \sim \,  \frac {T^5}{M_P^2 M_*^2} \,.
\label{thermalrate}
\end{equation}
As long as this rate is less than the Hubble expansion rate
(\ref{hubblerate}),  the observable sector cools down
predominantly due to the Universe's expansion, and  cosmology is
normal.  Thus, the normalcy temperature is defined from the
condition. $\Gamma\,  = \,  H$, which implies
\begin{equation}
T_C\, \sim \, (M_P M_W^2)^{1/3}\, \sim \,  10^8~GeV \, .
\end{equation}
This temperature is so high that it is completely safe to assume
that the other sectors have no visible effects at temperatures
much above BBN. In fact, since the permutation symmetric SM is an
effective theory valid only up to $M_*\, \sim \, TeV$, no
temperatures above this value can be considered within our
framework and any such discussion should await the UV completion
of the theory. Notice that below $M_*$ temperatures the rate
(\ref{thermalrate}) is even less than the rate of normal graviton
emission.  Already this fact indicates that such processes can be
safely ignored within the validity of our effective field theory
description.  Therefore we conclude that below the cutoff
temperatures, the thermal production of other copies via graviton
exchange is negligible,  and cosmology is standard.

From the above it also follows as no surprise that astrophysical
processes such as cooling of stars due to  graviton-mediated
emissions of hidden species give no constraints on the model.
Since all the known astrophysical objects are much cooler than
TeV, the inclusive rate of production of the hidden species within
such objects  (which is again given by (\ref{thermalrate})) is
much less than the graviton emission rate and is totally
negligible. This should be compared with the analogous result in the case of
large extra dimensions. In that case the production of light
Kaluza-Klein gravitons by supernovae provides one of the strongest
constraints on the model. Finally, we wish to note, that because of the presence of
$10^{32}$ exactly massless photons, the evapotation of astrophysical BHs is
$10^{32}$ times faster. This brings the mass of the smallest primordial BHs
that could survive till today, to approximately $10^{26}$g.

\subsection{Constraints from  New Gravitational States}
\label{newstates}

We now turn to the new states that must appear at the cut-off of our effective
description (assuming that a field theoretic description is
possible for such states). While the coupling between the species
induced by the graviton is totally negligible, one might wonder
whether the effective operators induced by the new gravitational
states at the cut-off are more dangerous phenomenologically.  Let
the new gravitational degrees of freedom that modify gravity
around the scale $M_*$ be $h_{\alpha}$.  Contrary to the graviton
coupling, the coupling of these states is not uniquely determined.
However,  the full permutation symmetry greatly restricts the number of
possibilities. Since these states should also obey the bound
(\ref{bound}),  their number cannot exceed $N$.  We can classify
these gravitational species by their transformation properties
under the permutation group.

If the gravitational states at the cut-off are singlets of the
symmetry $P(N)$  (just as the graviton is), then by permutation
symmetry the coupling of such gravitational species to the
different copies of the SM has the following form,
\begin{equation}
\label{coupling}
h^{(\alpha)} \, \lambda_{\alpha}\,  \sum_j \, \bar{\Phi}_j\, \Phi_j \, ,
\end{equation}
where the parameter $\lambda_{\alpha}$  defines the form and the
strength of the coupling. The spin of the $h^{(\alpha)}$ states is
not specified.  So the  $\lambda_{\alpha}$-symbols represent operators rather
than simple coupling constants.  For example, they can depend on
the derivatives that act on the different fields entering the
vertex. The total emission rate of $h_{\alpha}$ in SM collision
processes is given by,
\begin{equation}
\Gamma_{SM \rightarrow h} \, \sim \, \sum_{\alpha} \, E \, |\lambda_{\alpha}(E)|^2 \,
\end{equation}
where the summation is over the energetically available states.
Unitarity implies that for $E \lesssim  M_*$,
\begin{equation}
 \sum_{\alpha}  |\lambda_{\alpha}(E)|^2 \,  \lesssim \,  1 \,.
\end{equation}
The direct emission of gravitational species becomes relevant only
at the energies comparable to $M_*$, but their virtual exchange
can generate effective interactions between the different SM
copies. For example, the effective four-fermion couplings among
the  fermions of $i$-th and $j$-th copies can be  generated  via
exchange of $h^{(\alpha)}$ species as in Fig. \ref{cross}, and at
low energies will have  the following form,
\begin{equation}
\bar{\Psi}_i \Psi_i \bar{\Psi}_j\Psi_j \, \sum_{\alpha} \, {\lambda_{\alpha}\lambda^*_{\alpha} \over  M_{\alpha}^2}  \, .
\end{equation}
Here $M_{\alpha} \, \sim \, M_*$ are the masses of the
gravitational species. Through this effective operator the
particles of the SM  can annihilate into the hidden sectors. The
total rate of annihilation at energy $E$ goes as,
\begin{equation}
\Gamma_{SM \rightarrow hidden~SM} \, \sim \, E^5\, \left ( \sum_{\alpha} \, {\lambda_{\alpha}\lambda^*_{\alpha} \over  M_{\alpha}^2}  \right ) ^2  \, N\, ,
\end{equation}
where the factor of $N$ comes from the summation over all the
final copies. Low energy unitarity below the scale $M_*$ again
guarantees that  the above rate is compatible with  all the
existing phenomenological bounds. This is because the above rate
is bounded by  $E$  at any energy below the cutoff $M_*$. Thus,
for $E \lesssim M_*$,  we have $\Gamma_{SM \rightarrow
hidden~SM}\, \lesssim \,  E $. In the worst case scenario in which
$\lambda_{\alpha}$-s do not scale with energy and we neglect
possible cancellations,  we obtain that the rate scales as
\begin{equation}
\label{ratemax}
\Gamma_{SM \rightarrow hidden~SM}\, \lesssim \,  E^5/M_*^4 \, .
\end{equation}
In reality, for any sensible gravitational theory, in which
$h_{\alpha}$ couple to the conserved sources, the scaling of the
rate will be much more rapid.  For example, for the states coupled
to the energy momentum tensors of the SM species,
$\lambda_{\alpha}$-s scale as $E$, and correspondingly $\Gamma_{SM
\rightarrow hidden~SM}\, \lesssim \,  E^9/M_*^8$. However, to make
our arguments stronger we can use the rate (\ref{ratemax}) as the
maximal possible rate. In fact, this maximal possible rate
coincides with the production rate of KK gravitons in high energy
processes in  theories with two large extra dimensions.   Although
this fact is completely accidental, we can apply  the results of
\cite{ADD3}, to derive phenomenological bounds, since the latter
analysis solely relied on the above rate.  We thus see, that
thanks to permutation symmetry and unitarity even without knowing
the precise details of underlying gravity theory, we are able to
show that the phenomenological bounds are never more severe than
large extra dimensional scenario.

The other option is that the gravitational states transform under
$P(N)$, i.e.  are interchanged along with the SM copies (other
representations are excluded by unitarity due to the growth of
their dimension with $N$).
In such a case, the gravitational states can be labeled by the same indexes as the given SM copies.
By unitarity each gravitational degree of freedom $h_i$ can only couple with
the $i$-th SM copy with the maximal (in units of $M_*$) strength while with all the other copies
the strength of the coupling must be $1/N$ suppressed.  The emission rate of all $j\neq i$
gravitational species, in the collision of particles of $i$-th SM copy, will go as,
\begin{equation}
\sum \Gamma_{i \rightarrow j\neq i} \, \sim \, EN^{-1}\, ,
\end{equation}
where $E$ is the energy in the collision, assumed to be above the
mass of the gravitational species. This rate is negligible, and
the total  emission rate will be dominated by the emission of the
gravitational species ``belonging'' to the  same  SM copy. The
rate at energies above their mass goes as  $\sim
E^{n+1}/M_*^{n}$, where $n$ is determined by the type of the
interaction.  For example, for the emission of the spin-2 state
that couples to a conversed energy-momentum tensor, $n=2$.  Since,
in the case of exact $P(N)$ symmetry, change of gravitational laws
start  around the scale $\sim M_*$, the  mass of the corresponding
gravitational species responsible for this change is of order
$M_*$.  Hence, their  emission  becomes important only close to
the cutoff scale, which puts the phenomenological bound around
$M_* \sim$TeV.  A similar bound comes from the effective
high-dimensional operators generated by the exchange of the
gravitational species.  Notice that from this point of view the
phenomenology of gravitational species is very similar to the one
of the smallest BHs. This is not surprising, since as we have
argued above,  the smallest BHs are quantum objects, and
distinction between them and particles is rather blurry. In
certain sense, the smallest BHs themselves can be treated as new
gravitational species that are necessary for modifying laws of
gravity at the scale  $M_*$.

\subsection{Baryon Number Violation}

In theories with low gravitational cutoff, one of the most
pressing phenomenological questions is the strength of the baryon
number violating operators.  The source of such operators can be
virtual BH exchange. For example, consider a process in which two
$u$-quarks collapse into a virtual BH, which then evaporates into
an anti-$d$ and a positron,
\begin{equation}
 u\, + \, u\, \rightarrow \, BH \, \rightarrow \, \, d_c \, +  \, e_c \, .
\end{equation}
At low energies, below the BH mass this results into an effective baryon number violating operator of the form
\begin{equation}
uud_ce_c {f_{uud_ce_c} \over M_{BH}^2} \,.
\end{equation}
Here, we have parameterized the strength of the effective operator
by  a form-factor $f_{uud_ce_c}$ which encodes the probability
that a BH produced by two $u$-quarks can evaporate into $d_c$ and
$e_c$. The problem in drawing definite phenomenological
conclusions is due to the fact that not much is known about this
form factor for the smallest BHs, which are the most relevant ones
for the above process. For the large and heavy BHs, which decay
universally into all the species, this form factor is tremendously
suppressed, because the heavy BHs are classical.  As we have
argued above,  the smallest BHs on the other hand are not
democratic in species. All the potential problems for baryon
number violating operators, come from the fact that the democracy
is blindly assumed for the micro BHs. Although, we know, that for
the coupling with large number of species, the unitarity arguments
indicate precisely the opposite, unfortunately, for small number
of species we cannot prove the suppression of the non-diagonal
interactions rigorously.

What is the implication of the  above analysis for the current
framework of the $N$-copies of SM? First, here we can be sure that
the baryon number violating (or any other type) of
copy-non-diagonal processes are suppressed by powers of $N$.
However, for the processes inside each copy, we do not have such a
powerful statement. Therefore, in the worst case scenario, we
shall assume that unless additional measures are taken, for the
smallest BHs the form-factors are order one.  Then the small
virtual BHs will mediate the baryon number violating processes at
an unacceptable rate. The additional measure that we shall invoke,
is the {\it common} gauging of the baryon number symmetry of all
the different copies. That is, we shall postulate that the
$U(1)_B$ symmetry,  that corresponds to  the simultaneous baryon
number rotations of all the SM copies,  is gauged. The corresponding
gauge boson, $B_{\mu}$, we shall call the bary-photon. By
unitarity and $P(N)$ symmetry, the  $U(1)_B$  gauge coupling $g_B$
is suppressed by $1/N$ and thus, is of the gravitational strength,
$g_B \, \sim \, M_*/ M_P$.

The question that arises immediately in the case of gauging the
baryon number symmetry, is the anomaly cancelation.  In the
present case, the anomalies generate a gauge-non-invariant shift
of the Lagrangian density
\begin{equation}
q_W \, \alpha \sum_j \, {\rm Tr} (\tilde{F}\,  F)_j   \, ,
\end{equation}
where $\alpha$ is the gauge shift parameter,  $B_{\mu}  \, \rightarrow  \,  {1 \over g_B } \partial_{\mu}\alpha$,
where $j$ labels the dual $SU(2)$-field strength of different copies and
$q_W$ is the usual bary-weak mixed anomaly coefficient, which by permutation symmetry is common for all the copies. Similarly, there is also a gravitational anomaly and a mixed gravitational anomaly (see \cite{gravianomaly}),
\begin{equation}
q_G \, \alpha  \,  \tilde{R} \, R \,.
\end{equation}
In order to cancel the anomalies, we can implement a Green-Schwarz type
mechanism \cite{4dgs}. We shall achieve this, by introducing the
bary-axion, $b$, with the respective gauge and gravitational
couplings,
 \begin{equation}
{b \over M_b} \, \left ( q_W \, \sum_j \, {\rm Tr} (\tilde{F}\,  F)_j   \, + \,   q_G  \,  \tilde{R} \, R  \,\right ) \, ,
\end{equation}
that shifts appropriately under the baryon number symmetry,
 \begin{equation}
{b \over M_b} \, \rightarrow   \, {b \over M_b}  \, - \, \alpha \,
\end{equation}
As usual, for simultaneous  cancellation of all the anomalies the
coefficients $q_G$ and $q_B$ must be equal. This can be always
arranged by introducing extra fermions with the appropriate baryon
number and no charges under the SM gauge fields.  Again, by
unitarity the axion decay constant must be  $1/N$-suppressed and
thus be  gravitational,  $M_b \sim M_P$.

Anomaly cancelation by GS mechanism automatically implies that
the bary-photon  is a massive (Proca)  field, with  the bary-axion
being its longitudinal component  $B_{\mu} \, \rightarrow \,
B_{\mu} \, + \,  {\partial_{\mu} b \over g_B\, M_b} $. The mass of
the bary-photon is $m_B \, = \, g_B\, M_b \, \sim \, M_*$.  The
gauging of the baryon number forbids all the local baryon number
violating operators that could lead to the proton decay.  The
non-local ones, however, can be written down. These  are of the
following form,
\begin{equation}
\label{nonlocal1}
{\rm e}^{-\, i\, c\, g_B{\partial_{\mu}B^{\mu} \over \square} }\,  uud_ce_c  \, {1 \over M^2} \,,
\end{equation}
where $c$ is the total $U(1)_B$  charge of the fermions entering
in the operator. Equivalently, we can have the following operator,
\begin{equation}
\label{nonlocal2}
{\rm e}^{-\, i\, c \, {b \over M_b} }\,  uud_ce_c  \, {1 \over M^2} \, \,.
\end{equation}
Notice, that if it were possible to identify $b$ with the phase of
a  local scalar field, then the operator of the above form could
have been written in a local way. This however is impossible,
since going above the vacuum expectation value of the scalar
field, we should have recovered an anomaly free theory even
without $b$, which would contradict to the existence of the
anomaly in the baryonic current. So the above operators cannot be
generated by the exchange of the smallest virtual BH of mass $\sim
M_*$, since effective operators generated in this way must be
local.  At best, such operators must pay the price of
non-perturbative suppression, by the exponential factors exp$({1
\over g_B^2})$ and is probably negligible.

\section{Neutron Oscillations}

Gauging of the diagonal baryon number still allows mixing between
neutral states of different species. One phenomenologically relevant
possibility is the mixing of neutrons. Indeed,  gauge symmetries
allow the generation of the following  dimension-9 operators
\begin{equation}
\label{nn}
\kappa_{ij}\,{u_i\, d_i \, d_i \,  \bar{u}_j\, \bar{d}_j\, \bar{d} _j \over M_*^5} \,      \,
\end{equation}
where $\kappa_{ij}$ parameterize the strength of these operators.
The structure of the matrix $\kappa_{ij}$ is determined by the permutation symmetry.  For any
system of $N$  SM copies invariant under the full $P(N)$ group,  we have $\kappa_{ij} = \kappa$.

The above operator induces a mixing between neutrons of the different
SM copies
\begin{equation}
\label{nmixing}
\delta m_n \sum_{i\neq j} \,  \,  n_i \, n_j \, ,
\end{equation}
with
\begin{equation}
\label{deltam}
\delta m_n \, \sim \, \, 10^{-4} \, m_n \, \kappa\ \left ({m_n \over M_*}\right )^{5} \, ,
\end{equation}
where $m_n$ is the neutron mass, and the extra suppression factor
of order $10^{-4}$ comes from the three-quark-neutron  matrix
elements. The constraints on the parameter $\kappa$ are the
following.

The requirement of the three-level unitarity  of the quark scattering
processes below the scale $M_*$,  gives the constraint  $\kappa \, \lsim \, {1 \over
\sqrt{N}}$. However, taking into account loops with
intermediate quarks from different species, the constraint
gets more  severe.  Summing over intermediate species, the loop
expansion goes as series in  $N\kappa$, and this demands  $\kappa
\, \sim  \, 1/N$. Notice, that the latter value is compatible with
the expected suppression of the inter-species operators  generated
by the virtual BH exchange. We shall, therefore, adopt  this
value.

This mixing can lead to the transition of our neutron into the
hidden ones. Such a mixing with a single mirror copy
was already studied in \cite{berezhiani} and as we shall see the case
with many copies introduces some qualitatively distinct features. In the case
of many species an analog mixing for the Higgs was considered in \cite{foot2}.

Of course, due to exact degeneracy of neutron masses,  the transition is normally
forbidden in nuclei due to energy conservation (a neutron that oscillates in an
hidden copy would not be bound and so would have
higher energy), and therefore will not be in conflict with the usual neutron
disappearance bounds inside nuclei \cite{neutronbound}. But this
would not apply to free neutrons which could indeed oscillate
into the hidden ones. We shall see, that the transitions
can take place and lead to some potentially  observable
effects.

\subsection{Vacuum Oscillations}

We now wish to give a detailed description of this effect. By taking
into the account the mixing between the neutrons of the different SM copies,
the full $N\times N$ neutron mass-matrix takes the following form (see also \cite{foot2}),
\begin{eqnarray}
\label{neutronmass}
m_{ij} = \left(
\begin{array}{cccc}
m_n & \delta m_n & \delta m_n &..\\
\delta m_n & m_n & \delta m_n & ..\\
\delta m_n & \delta m_n & m_n & .. \\
.. & .. & .. & .. \\
\end{array}
\right)
\end{eqnarray}
Notice, that since due to exact permutation symmetry all the
off-diagonal entries are equal, this matrix can be written in a
very simple form
\begin{equation}
\label{neutronmass1}
(m_n\,- \delta m_n) \sum_{j}\, n_j\,  n_j\, + \, \delta m_n \, \left(\sum_{j}\, n_j \right)^2 \, .
\end{equation}
We can rewrite the above as,
\begin{equation}
\label{neutronmass2}
m_n \, n_1 \, n_1 \, +  \,
\delta m_n \,  \left (n_1 \,\sum_{j \neq 1} n_j+n_1 \,\sum_{j \neq 1} n_j \right )\, + \, \delta m_n \,
\left (\sum_{j\neq 1}\, n_j \right)^2 \, + \,
(m_n\,- \delta m_n) \sum_{j \neq 1} n_j\,  n_j\, .
\end{equation}
where $n_1$ denotes the neutron from our copy.
Introducing the notation $n_h \, = \, {1\over \sqrt{N-1}}\, \sum_{j\neq 1}\, n_j$,
the problem reduces to the following $2\times 2$ mixing between the two states $n_1$ and $n_h$,
\begin{eqnarray}
\label{lightheavy}
 \left(
\begin{array}{cccc}
m_n & \delta m_n \, \sqrt{N-1}\\
\delta m_n \, \sqrt{N-1} & m_n\, + \, \delta m_n \, (N-2)\\
\end{array}
\right)
\end{eqnarray}
In addition there are $N-2$ exactly degenerate orthogonal states
with mass $m_n-\delta m$, which do not  mix. The above matrix  has
two mass eigenstates: $n_H \, \equiv \, {1 \over \sqrt{N}} n_1 \,
+ \, {\sqrt{N-1} \over \sqrt{N}}\, n_h$ and $n_1' \, \equiv \, -
{1 \over \sqrt{N}} n_h \, + \, {\sqrt{N-1} \over \sqrt{N}} \,
n_1$, with masses $m_H\, = \, m_n \, + \, (N-1)\delta m_n \, $ and
$m_1' \, = \, m_n-\delta m_n$ respectively. Thus, in total there are
$N-1$ exactly degenerate states of mass $m_n-\delta m_n$, and a
single state of mass $m_H$.  In the other words, the mass
eigenstates decompose into the $N-1$ dimensional representation of
the permutation group $P(N)$ plus a singlet. Each particular
neutron mixes with a single common state, in which all the other
neutrons enter democratically. Thus, our neutron will oscillate
into the other neutrons through $n_H$- state, and evolve in time
as
\begin{equation}
\label{osc}
n_1(t)  \, = \sqrt{{N-1 \over  N}} \,n_1' \, + \, {1 \over \sqrt{N}} \, n_H \, {\rm e}^{-i\,N\,\delta m_n  \, t} \, .
\end{equation}
The inclusive probability of disappearance is easily obtained to be:
\begin{equation}
P(t)\approx \frac 4 N \sin^2 \left (\frac {N\, \delta m_n\, t}2  \right ) \,.
\end{equation}
The frequency of the oscillations is $1/\tau \, \sim \, \delta m_n\,  N$, and can be
quite large. Using (\ref{deltam}) and setting $\kappa \sim 1/N$,  we get
$1/\tau \, \sim \, 10^{-10} ({\rm TeV}/M_*)^5$ eV, implying  the
oscillation period of approximately
\begin{equation}
\label{periodn}
\tau \, \sim \, 10^{-5}
\left (M_* \over {\rm TeV} \right )^5 {\rm sec} \,.
\end{equation}
However the important point is that the amplitude
of the oscillation is  suppressed by $1 \over \sqrt{N}$,
independently of  $\delta m_n$. This gives the total disappearance
probability of neutrons (in our SM copy) in the vacuum  $P_{dis} \, \approx \, 2/N$.

\subsection{Oscillations in the Magnetic Field}

Typically  our neutrons do not propagate in the vacuum,
and this fact can change the transition  probability dramatically.
In particular,  the presence of the magnetic field can trigger  a
resonant transition. The previous analysis can be readily
generalized in the presence of an additional potential $\epsilon$
(e.g., due to magnetic field) for our neutron $n_1$. Since the
potential is only experienced by our neutron, its presence affects
only $2\times 2$ mixing matrix (\ref{lightheavy})  of $n_1$  and  $n_h$ states, which
in this case becomes,
\begin{eqnarray}
\label{lightheavy1}
m_{ij} = \left(
\begin{array}{cccc}
m_n\, + \epsilon & \delta m_n \, \sqrt{N-1}\\
\delta m_n \, \sqrt{N-1} & m_n\, + \, \delta m_n \, (N-2)\\
\end{array}
\right)
\end{eqnarray}
The presence of the extra potential can resonantly
increase the oscillation amplitude if the two states become nearly
degenerate. This happens when
\begin{equation}
\label{epsilonr}
\epsilon \, \approx \, \delta m_n
\, (N-2) \, .
\end{equation}
In such a case, the mixing angle becomes $45^o$ and
the resonant oscillation period is
\begin{equation}
\label{periodr}
\tau_r \, \approx \, (\delta m_n
\,  \sqrt{N-1})^{-1}.
\end{equation}
In particular, the source of the potential
can be a magnetic field $B$.  In this case $\epsilon \, = \,
\vec{\mu}. \vec{B}$, where $\mu$ is the neutron magnetic moment.
We are now ready to consider some phenomenologically interesting regimes.

\subsection{Different Regimes}

Let us note, that since the  transition amplitude in the non-resonant case is suppressed by
 $1/\sqrt{N}$, the phenomenologically most interesting situation is when $N$ is not  extremely large. In the present context this implies that the symmetry should be less than the full permutation group $P(10^{32})$, in such a way that each neutron  has a permutation symmetric mixing with the subset of the copies, less than the total number.  This smaller subset then will define a set of ``nearest neighbors'' in the space of species.  In such a case, the number  $N$ in our previous calculations has to be understood as the number of such neighbors, as opposed to the total number of species,  which we shall temporarily denote by $N_{TOTAL}\, \sim \, 10^{32}$.   The symmetry justification for such a situation is straightforward.
We can split all $N_{TOTAL}$ SM copies into $n=N_{TOTAL}/N$ subgroups,
with $N \, \ll \, N_{TOTAL}$ members in each subgroup. We then require
the full permutation symmetry within each group, plus symmetry of
all possible permutations among the groups. The resulting symmetry
group thus is, $P(n)\, \times \, \left (P(N)_1 \, \times
P(N)_2 \, \times \, ...\, P(N)_n \right )$, where it is
understood that the elements of $P(n)$ exchange groups among each
other. Requiring such a symmetry, the neutron mass
matrix splits into the $N\times N$ blocks. The blocks on the
diagonal have the form (\ref{neutronmass}).  The mixing entries in these blocks
are much larger, since the six-quark operators (\ref{nn}) that mix
neutrons from the same group,  are only restricted to be
suppressed by $\kappa \sim 1/N$.  All the other entries of the
full $N_{TOTAL}\times N_{TOTAL}$ matrix,  that mix neutrons from different
subgroups, as before, must be suppressed by $\kappa \sim \, 1/N_{TOTAL}$,
and are negligible.  Therefore, for each given neutron, the
influence of all the other subgroups can be ignored, and the
problem reduces to $N\times N$ mixing. Thus,
we can simply use all our previous results, keeping in mind that
$N$ is now not the total number of copies but only of our close neighbors.

Having freed the parameter $N$, we can now discuss some experimentally interesting regimes of neutron oscillation.

\subsubsection{Negligible Magnetic Field}

The first regime takes place when the neutron potential created by the magnetic field is negligible with respect the mass difference,  $ \epsilon \, \ll \, \delta m_n \, (N_q\, - \,2)\,$.   In this case oscillations proceed as in the vacuum, and the oscillation period is given by  (\ref{periodn}).  So  in any measurement performed with a time resolution  $\Delta t \gg \tau$, the neutron will appear in the hidden state with probability  $2/N$.  In experimental measurements this will manifest itself in form of the unaccounted neutrons.

The relevant experiments are the ones on neutron lifetime measurements \cite{lifetime} and the cold neutron experiments that directly test oscillations of neutrons into the the mirror copies \cite{neutrondisapp1, neutrondisapp2}.  In order to place the constraints on our parameters the experimental results must be compared with the present model.
For example, let us consider the effect of the above oscillations  on the measurement of the  neutron lifetime.  The neutron decay can be easily accounted by giving an imaginary part $i/\tau_n$ to the diagonal Hamiltonian. Since the copies of SM are identical, the decay rates of all neutron into their own SM species are the same. In the usual case of $N=1$, this leads to the familiar decay law,
\begin{equation}
N_n(t) \, = \,  N_n(0) e^{-t/\tau_n} \, ,
\end{equation}
where $N_n$ is the number of neutrons. However, for many copies,
the law changes. In particular in this case when
$\tau \ll \tau_n$ the number of neutrons gets modulated by the
rapid oscillations,
\begin{equation}
N_n(t) = N_n(0) \left(1 \, - \, \frac 4 N \sin^2 \left (\frac {t}{\tau}  \right )\right)
e^{-t/\tau_n} \,.
\end{equation}
For the measurements produced with the time resolution  $\Delta t \gg \tau$, one sees an average value of the pre-factor
\begin{equation}
N_n(t) = N_n(0) \left (1 \, - \, {2 \over N} \right ) \,
e^{-t/\tau_n} \,
\end{equation}
Thus, the modulation cannot be captured in the measurements that are sensitive only to relative decrease of the neutron number,  if the time resolution of the measurement is  $\Delta t \gg \tau$.
In such  a case in order  to place the bound on $N$,  the mechanisms responsable for systematic neutron losses as well as the impact of the neutron-detection measurement on the oscillation process have to be well understood.

\subsubsection{Resonant Oscillations}

The second interesting regime is the resonant transition in the magnetic field.  As said above,  this happens when the condition (\ref{epsilonr}) is satisfied.   In such a case, the mixing angle becomes $45^o$ and the resonant oscillation period is given by,
\begin{equation}
\label{periodr}
\tau_r \, \sim \, 10^{-5}\, \sqrt{N-1} \, \left (M_* \over {\rm TeV} \right )^5 {\rm sec} \,.
\end{equation}
For example, for  the earth magnetic field, $.5~ G$, one has $\epsilon \, \sim \,
10^{-12}$eV.  For $N > 2$ and $M_* =$TeV, the resonance happens in a strong magnetic field $\sim 10^2$G or so.

However,  both the resonant value of the magnetic field as well as the oscillation period are very sensitive to the value of $M_*$ in  TeV units.  So  taking $M_*$ only few TeV  decreases the  resonant value  of the magnetic field, and increases the oscillation time dramatically. To illustrate the point, notice that for $M_* \, = \, 5$ TeV or so, the value of the resonant magnetic field would be the one of the experiments \cite{neutrondisapp1, neutrondisapp2 }. The lower bound on the oscillation time derived from these experiments ($\tau > 103$ sec from \cite{neutrondisapp1} and  $\tau > 448$ sec from \cite{neutrondisapp2} respectively), would translate into the bound on the number of neighboring copies  of approximately $N \, \gtrsim 10^{9}$.  From the above consideration it is clear that for testing this framework, one has to scan entire portion of the parameter space corresponding to the values of  $M_*$ within the inteval of $1-10$~TeV, dictated by the hierarchy problem.  For this purpose,  performing  the experiments of this type for the different values of the magnetic field would be extremely important \cite{neutronexperiment}.  In such a case, in the experiment with ultra cold neutrons, one should see a resonant increase of the missing neutron number for certain values of the magnetic field.

\section{Neutrinos and Dark Matter}
\label{bsm}

Experimentally the two most concrete indications that the SM is
incomplete are provided by the presence of dark matter in the
Universe and by neutrino masses. In this section we wish to show that our framework
offers a novel way of generating the small neutrino masses as well as
possible dark matter candidates.

\subsection{Small Neutrino Masses from Large Number of Species}

Observation of neutrino oscillations has shown that neutrinos have
masses of the order $10^{-1}~eV$ (for a review see
\cite{strumia}). In trying to understand the underlying physics
responsible for the small neutrino mass, one can think in terms of
lepton number symmetry. If lepton number is only an
approximate symmetry of the low energy world and is not respected
by  physics at some high scale, the small neutrino mass can arise
as a result of the lepton number violating effective operator
suppressed by the cutoff scale. In such a case, neutrinos can be effectively
Majorana particles. This scenario is explicitly realized
by  the seesaw mechanism \cite{seesaw}. In this case, the lepton
number is maximally violated by a large mass  of the right-handed
neutrino. Since this state carries no gauge charge under the SM
group,   it is not prevented by the latter gauge symmetry  from
having an arbitrarily high mass. After integrating out this heavy
fermion, one is  left with the small  Majorana mass of the SM neutrino.

The  large extra dimensions scenario, offers an alternative
explanation for the smallness of the neutrino mass \cite{neutrinosextra}. In contrast
to the  see-saw,  in the minimal version,  the
lepton number is assumed to be a good symmetry of the low energy
world  (this can be achieved, e.g., by gauging $B-L$ symmetry). In
this case, neutrino is bounded to be a Dirac particle.  The
smallness of its mass is then  guaranteed by the fact that the
right-handed neutrino, very much like the graviton, is delocalized and
lives in the extra dimensional bulk.  For this reason, the extra-dimensional
neutrino  is forced to be extremely weakly coupled to its left-handed gauge
non-singlet partner. Notice, that although the two scenarios are
dramatically different, they both employ the crucial property that
the right-handed neutrino is a SM gauge singlet. It is this very
property that allows the right-handed state to either be very
heavy or to be  freely spread-out into the extra dimensional bulk.

We now wish to suggest an alternative mechanism for generating the
small neutrino mass,  in the framework of large number of the SM
copies. Just like in the extra dimensional scenarios, the
new mechanism crucially relies on the fact that the
right-handed neutrino is a gauge singlet, and  like the
graviton can share its couplings with the different SM copies. As
a result of this democracy, the couplings and the resulting
neutrino masses are suppressed by  $1/\sqrt{N}$ factor.

For simplicity of illustration, we present our mechanism  for the case of single lepton generation in each SM copy.
Generalization to three generations of neutrinos, is trivial.
In order to generate neutrino masses we assume that each copy of
the SM is endowed with a right handed neutrino $\nu_{Rj}$, which
we shall label by the same index $j$. However, since right-handed
neutrinos share no SM gauge charges,  the notion of ``belonging''
to a given copy is only defined in terms of transformation
properties with respect to the permutation group. The
right-handed neutrino sharing the same label with the given SM
copy is defined as the one that under $P(N)$ permutation symmetry
transforms simultaneously with the rest of the particles belonging
to this copy.

As for the neutrons, the right handed neutrinos do not have
SM charges so they can mix with states in the other sectors. In a
sense the neutrinos live in the ``bulk'' of the space of species
just as the graviton does.  With each SM copy the right-handed
neutrinos interact through the dimension-4 gauge-invariant
operators,
\begin{equation}
(HL)_i \lambda_{i j} \nu_{Rj}\, + \, h.c.
\label{neutrinocoupling} \,
\end{equation}
where $\lambda_{ij}$ are Yukawa coupling constants, that represent
$N\times N$ matrix in the space of species. As said above,  we
assume that no right-handed neutrino Majorana mass arises ( as a
consequence,  for example,  of  $B \, - \, L$ number conservation).

The form of the Yukawa coupling matrix between the different species is
restricted by the permutation symmetry to be of the following form,
\begin{eqnarray}
\lambda_{ij} = \left(
\begin{array}{cccc}
a & b & b &..\\
b & a & b & ..\\
b & b & a & .. \\
.. & .. & .. & .. \\
\end{array}
\right)
\end{eqnarray}
After taking into the account the non-zero expectation values of
the  SM Higgs fields, the above Yukawa matrix translates into the
mass matrix for neutrinos. Here we make the minimal assumption
that the permutation symmetry is not  broken by the Higgs VEVs,
and thus, they are all equal $\langle H_j \rangle \, = \,  v$.
Then, the mass matrix is $m_{ij} \, = \, \lambda_{ij} \, v$. This
is very similar to the one of neutrons (\ref{neutronmass}). However, since unlike neutrons,  the
right-handed neutrinos are introduced as elementary fermions, the
diagonal and non-diagonal entrees have no reason to be as
different, as they were in the case of neutrons. So, it is natural
to assume that, to the leading order in $1/N$, the coefficients
$a$ and $b$ are roughly similar.

In order to find the resulting neutrino masses, notice that the
coefficient $b$ cannot be large because of unitarity.  For
example, consider a process of right-handed neutrino production in
the scattering of the SM states.
The inclusive rate for this process goes as,
\begin{equation}
\Gamma_{tot}\approx N \, b^2 \, E \, ,
\end{equation}
where the factor $N$ comes from the summation over the different $\nu_{Rj}$ species in the final state.
Unitarity below the cutoff scale then demands,
\begin{equation}
b \le \frac  1 {\sqrt{N}} \,.
\end{equation}
On the other hand the coefficient $a$, the coupling of a the right
handed neutrino to its ``own'' specie is not so restricted by unitarity
and could be taken to be larger. However, the two couplings are exactly
the same in nature, we  shall not require any strong hierarchy between them.

We are now ready to compute the neutrino masses. When the Higgs in
each sector acquires a VEV,  from (\ref{neutrinocoupling}) one
obtains Dirac masses for the neutrinos. The structure of the
mixing matrix is identical to the one of neutrons, studied  in the
previous section in detail,  so we can just use the results
there. The mass eigenvalues are given by
\begin{eqnarray}
m_i&=& (a-b) \, v~~~~~~~~~~~i=1,...,N-1 \\
m_N&=& [a+(N-1)b] \, v
\end{eqnarray}
As a consequence of the permutation symmetry all neutrinos are
degenerate while one has a mass of the order of the Planck scale.
The hierarchy between the heavy and the light neutrino masses is
$\sim \, N$. Notice, that the latter fact  can be considered as an
independent  (from unitarity) reason for  the bound $b\,  \lesssim
\, 1/\sqrt{N}$. In the opposite case, effectively we would be
forced to introduce an elementary particle heavier than the Planck
mass, which would make a little sense.  Hence, in our approach,
we can equally say that the SM neutrinos are forced to be light in order
to avoid the appearance of the state with the  super-Planckian
mass! The latter argument  bares no reference to unitarity.

Assuming that $b\approx 1/\sqrt{N}$ the neutrino masses
in our scenario are
\begin{equation}
m_\nu \approx (a/b \, - \, 1)\frac {v}{\sqrt{N}} \, \sim \,  (a/b \, - \, 1) {vM_* \over M_P}
\end{equation}
Depending on the precise value of the $M_*$, the right magnitude of neutrino masses
requires a mild hierarchy $a\approx 1000 \,  b$.
Notice that, the above small hierarchy between $a$ and $b$ corresponds to a right-handed neutrinos being somewhat ``localized'' in the space of species.

One might be worried that due to the large number of light
neutrino eigenstates there might be very large oscillations into
the hidden neutrinos which would be inconsistent with observation.
This however does not happen as can be seen by simply
adopting the results of the neutron oscillations derived earlier.
Exactly by the same reasons as in neutron case,  any given
neutrino (call it $\nu_1$) oscillates into a single common  heavy
state  $\nu_H$, according to the following rule,
\begin{equation}
\label{oscnu}
\nu_1 \, \approx \left(1-\frac 1 {2N}\right) \,\nu_1' \, + \, {1 \over \sqrt{N}} \, \nu_H \, {\rm e}^{-i\,N\, b\, v  \, t} \, .
\end{equation}
The only difference from the neutron case  is that the state
$\nu_H$ is ultra heavy,  so there cannot be oscillations into the
neutrinos of the hidden sectors. The oscillations will be purely
between neutrinos within the same SM copy, just as in the usual
case. This fact should be compared with the analog mechanism in large extra
dimensions \cite{neutrinosextra}. In that case the neutrinos can oscillate
into the states of the KK tower which behave as sterile
neutrinos. This needs to be suppressed  because such  oscillations
are disfavored by the present data \cite{strumia}.

Having introduced renormalizable couplings between different
copies one might wonder about other renormalizable operators
connecting the species. In fact there are only two operators of
this kind which are gauge invariant, the Higgs quartic coupling
and the mixing between the $U(1)_{hypercharge}$,
\begin{equation}
\lambda^H_{ij} |H|^2_i |H|^2_j~,~~~~~~~~~~~~~~~~\alpha_{ij}
F_{\mu\nu}^i F^{j \mu\nu}
\end{equation}
The kinetic mixing of photon with a hidden sector gauge field  is known to induce the effective charge-shifts \cite{Holdom:1985ag} (for a stringy realization see \cite{kineticmixing} and also \cite{abel} for a more phenomenological discussion).  This remains true in the present case also.
So the second coupling above  effectively gives a charge to the SM fermions
of specie $i$ under the gauge group $j$ (see \cite{rahman}).
For the Higgs quartic coupling (and similarly for $\alpha_{ij}$)
the model is perturbative when,
\begin{equation}
\lambda^H_{ij}\le \frac 1 {N}~~~~~~~~~~~~~~i \ne j
\end{equation}
and $\lambda^H_{ii} \le 1$ which coincides with the value demanded
by naturalness. This induces a coupling between different species
comparable to gravity and therefore is completely negligible for all practical
purposes.

\subsection{Dark Matter}

The fact that the cross section for processes between the
different species are so weak, makes the  baryons in the other
sectors into the viable dark matter candidates. In this section
we would like to discuss such a possibility. The key point that
can make the dark matter effectively collision-less is the large
number of copies with very low individual densities.

This idea was introduced in Ref. \cite{kaloper} in the context of
large extra dimensions. The authors considered a brane folded in
the extra-dimension proposing that dark matter could just be
ordinary matter from a different fold. Such a matter  is
electromagnetically invisible because a ray of light should travel
a distance larger than the horizon size in order to reach us. On
the other hand such matter has an usual gravitational effect on
the Universe since the four-dimensional  gravity is common. Hence
the baryons from the other folds are the natural candidates for
the dark matter.

This picture becomes natural in our framework due to the fact that  we do not
need to postulate extra dynamics that produces the extra copies, such as the folded
braneworld.  Moreover,  the huge number of copies automatically
takes care of the needed dark matter properties.
Cosmological observations require that the Universe is made of
$74\%$ dark energy, $~22\%$ dark matter and $~4\%$ visible
baryons. If only $n$ copies are populated, we need a total baryon
density from the SM copies which is about 5 times the one of
visible baryons,
\begin{equation}
n~n_B^{dark}\approx 5 n_B^{visible} \, .
\end{equation}
The minimal number is constrained by the fact that the star-formation should not
have started in the other copies or else this would be inconsistent with observations,
\begin{equation}
n^{dark}_B \leq \frac 1 4 \,.
\end{equation}
Close to the bound star formation would have just started in the
other copies and exotic objects such as MACHOS might have been
formed in the other sectors. However, when put in the
cosmological context, the actual number of the required
weakly-populated copies may be much higher.  The outcome depends
on the concrete scenario of how in the post inflationary Universe
the degeneracy between the copies is broken, and how the reheating
takes place. One necessary requirement is that our copy is the
maximally populated one.  On the other hand,  the degeneracy among
the other  copies may be lifted in a different ways.

A crucial question thus is how the asymmetric initial conditions
on the visible and dark matter abundance can be produced.   A
possible mechanism was introduced in \cite{Dvali:2009fw} where it
was shown that this can be obtained through the modulated
reheathing mechanism of Ref. \cite{modulated} within
low scale inflationary models. In this scenario the equal
under-population of all the hidden sectors is an automatic
consequence of the  density perturbation mechanism in which the
perturbations are imprinted during the reheating, via a common
modulating field. All the other sectors then have equal tiny
densities, and are never thermalized. The advantage of this
picture is that dark matter abundance is calculable in terms of
known cosmological parameters, such as baryon to entropy ratio and
density perturbations.  Very low individual densities also make
the outcome independent of the concrete mechanism of baryon
asymmetry generation,  since this mechanism is not  operative in
the hidden sectors anyway due to their very low densities. Thus,
regardless of this mechanism, other copies effectively contain
equal number of baryons and anti-baryons.

In general, other  scenarios are also possible, in which asymmetry
among the observable and visible sectors is more involved. For
example, we can have a situation in which not all the hidden
sectors are equally populated, and some are having densities
comparable to ours.  In such a case,  such sectors cannot include
an equal number of hidden baryons and anti-baryons, due to their
effective annihilation,  and the details of baryogenesis become
relevant.  In addition,  as we mentioned earlier, in order to
evade BBN constraints there should not be any extra massless
degrees of freedom which contribute to the expansion of the
Universe. This implies that the temperature of the other copies is
much smaller that the one measured in cosmic microwave background,
\begin{equation}
T_{dark}<< 3 K \, .
\end{equation}
This puts extra burden on the cases  when densities in the hidden
copies are relatively high, since in such a case the species there
have to be produced very cold. In general the cosmological
evolution will be determined by the matter density and temperature
of each SM copy, $n_B^i,~T^i$. and a  detailed analysis is needed.

Finally we wish to quote that some constraints of many species inflationary cosmology,
applicable to the regimes that differ from the ones of our interest above, were considered in \cite{Ncosmo}.

\section{Discussion}

In this paper we have investigated varius phenomenological aspects of the
Large $N$ species solution of the hierarchy problem proposed
recently, focusing on the permutation symmetric scenario obtained
by adding $10^{32}$ mirror copies of the Standard Model.

The key point in addressing the hierarchy problem is played by the intrinsic
connection between the number of species and short distance gravitational physics. According o this connection,  the  existence of $N$ species in the low energy world automatically lowers the gravitational cutoff (the quantum gravity scale) according to the eq. (\ref{bound}). In other words, the existence of species in the low energy theory implies extra-dimensional type modification of gravity at distances much larger than the Planck length.

We have shown, that some of the properties of this new gravitational physics can be constrained by symmetries acting on the low energy species (in our case SM copies).
In particular, cyclic symmetry among the copies, gives rise to gravitational physics
strikingly similar to the one of extra dimensions. The existence of  many SM copies,
however, makes phenomenology different from the large extra dimensional framework.
The maximal departure from the smooth geometric picture takes place in the limit of full
permutation symmetry acting among the copies. In this case the hidden copies behave as perfect dark worlds. Exact permutation symmetry combined with unitarity turns out to be a remarkably powerful principle so, not surprisingly, this version of the theory is maximally restrictive.

The basic result of our analysis is that when confronted with the existing cosmological and particle physics data below the TeV scale, the scenario easily passes all possible tests, but at the same time opens up new experimental avenues for the search of new physics.
One model-independent prediction is strong gravity around TeV energies.
This, among other things, should manifest itself in the production of microscopic BHs in particle collisions. However, the observable properties of these BHs  (such as the evaporation rate to mass ratio, or branching ratios into the visible versus invisible channels) are very different from the canonical
large extra dimensional picture.   For example, unlike the usual large extra dimensional black holes, the branching ratio for evaporation into the hidden channels increases with the BH mass according to (\ref{fraction}).

Secondly, there is an interesting  low-energy  window into the hidden world  in the form of the oscillations of the neutron into its hidden copies. The existence of many potential mixing-partners brings a qualitatively new experimentally interesting features, such as the possibility of rapid oscillations with a suppressed amplitude, or the resonant transition in a non-zero magnetic field.  This opens up an exciting new possibilities  of searching for an imprint of the dark sector  physics in cold neutron experiments of the type\cite{neutrondisapp1, neutrondisapp2}.

We have also shown that some of the mechanisms introduced in the
context of large extra dimensions can be generalized naturally to
the case of many species and do not rely on the existence of an
actual underlying geometry. In particular proton stability can be
guaranteed by the gauging of the common baryon number between the
species. (This still allows the above mentioned mixing between the neutral states from
different species, particularly neutrons). The neutrinos can acquire  a
small,  $1/\sqrt{N}$-suppressed,  Dirac mass without seesaw mechanism.
Baryons from the hidden copies are viable candidates for dark matter. It is
rather clear that many of the features studied in the context of
large extra dimensions do not rely on the existence of a geometry but on a more
general notion of unitarity and locality.

The most relevant open theoretical question to be understood is the UV
completion of this scenario. As we have seen in section \ref{newstates}
the permutation symmetry already restricts strongly the possible
physics at the cut-off.  In  the case of cyclic symmetry, which implies UV completion
in terms of smooth extra dimension type geometry,  the possibility of embedding in string
theory, along the lines of \cite{ADD2}, seems
not to exhibit any difficulty of principle. However,  UV completion of the full permutation
symmetric scenario  appears much more mysterious because the departure from geometry is maximal. Is there a stringy realization of this scenario? While we are not aware of any explicit construction it seems natural to look for this type
models in the non-geometrical compactifications of the landscape
of string theory (see for example \cite{nongeometric}). These constructions
may not contain K\"ahler
moduli describing the size of physical dimensions and so appear as
the appropriate starting point. In general, consistent propagation
of strings requires a two dimensional conformal field theory of the
appropriate central charge. In the standard case the conformal
field theory is a free theory and has as low energy limit 10D
supergravity. However it is possible to replace the 6-extra
dimensions with a conformal field theory of equal central charge
which does not admit a geometrical description at low energy. The large
number of light fields could be perhaps obtained by introducing
branes in this setup. Alternatively it might be possible to construct
our non geometrical setup considering compactifications with
complicated topology (see for example \cite{nflation}). In this
way it might be possible to obtain an exponential proliferation of
light modes which in turn creates a large hierarchy between the
Planck scale and the fundamental scale. We leave a detailed
investigation of these ideas to future work.

\vspace{1cm}
{\bf Note Added:}\\ Beside neutron oscillations mixing between other
neutral states might also be of interest experimentally. In particular it
was pointed out to us that measurements of positronium oscillation
have been performed \cite{positronium} which excluded oscillations
with a precision of $4 \times 10^{-7}$. Unfortunately in our case the
mixing between these states is strongly suppressed by the size of the
object so that we do not expect this to be in the experimentally interesting
region due to the long period of the oscillation. Another interesting window
might however appear in $K_0-\bar{K}_0$ oscillations.
Dimension 6 operators of the form,
\begin{equation}
\frac 1 N \frac {u_i \bar{s}_i\, \bar{u}_j s_j}{M_*^2}
\end{equation}
would induce oscillations between Kaons of different copies (we assume as for the
neutrons that mixing within the same copy are suppressed). With $M_*\approx$ TeV
the oscillation time, $\tau^{-1}= O(m_K^3/M_*^2)$, is much shorter than
the experimental bound from flavor physics. Given that the transition
$d \bar{s}\to s \bar{d}$ is measured with a 1\% precision this could
already imply a rough bound $N \approx 100$ on the minimum number of copies.

\vspace{1cm}
{\bf Acknowledgments}

We thank Diego Blas, Savas Dimopoulos, Hyung Do Kim, Gregory Gabadadze,
Holger Nielsen, Riccardo Rattazzi, Sergey Sybiriakov and Andrea Wulzer for very
useful discussions and comments. Special thanks are to  Yuri Kamushkov, Anatoly Serebrov and
Alexander Mtchedlishvili for illuminating  discussions on neutron experiments.
The work of G.D. is supported in part by David and Lucile Packard
Foundation Fellowship for Science and Engineering, by NSF
grant PHY-0245068 and by ERC grant on TeV physics.

\end{document}